\documentclass[preprint]{revtex4}

\usepackage{graphicx}

\usepackage{amssymb,amsfonts,amsmath,color}

\newcounter{nameOfYourChoice}

\begin{document}

\title{Stochasticity-induced stabilization in ecology and evolution: a new synthesis}
\author{Antony Dean$^{1,2}$ and Nadav M. Shnerb$^3$}

\affiliation{$^1$ Department of Ecology, Evolution, and Behavior, University of Minnesota, St. Paul, MN 55108 \\
\noindent $^2$ BioTechnology Institute, University of Minnesota, St. Paul, MN 55108 \\ $^3$ Department of Physics, Bar-Ilan University, Ramat Gan
52900. Israel. }

\begin{abstract}
The ability of random environmental variation to stabilize competitor coexistence was pointed out long ago and, in recent years, has received considerable attention. Analyses have focused on variations in the log-abundances of species, with mean logarithmic growth rates when rare, $\mathbb{E}[r]$, used as metrics for persistence. However, invasion probabilities and the times to extinction are not single-valued functions of $\mathbb{E}[r]$ and, in some cases, decrease as $\mathbb{E}[r]$ increases. Here, we present a synthesis of stochasticity-induced stabilization (SIS) phenomena based on the ratio between the expected arithmetic growth $\mu$ and its variance $g$. When the diffusion approximation holds, explicit formulas for invasion probabilities and persistence times are single valued, monotonic functions of $\mu/g$.  The storage effect in the lottery model, together with other well-known examples drawn from population genetics, microbiology and ecology (including discrete and continuous dynamics, with overlapping and non-overlapping generations), are placed together, reviewed, and explained within this new, transparent theoretical framework. We also clarify the relationships between life-history strategies and SIS, and study the dynamics of extinction when SIS fails.
\end{abstract}

\maketitle

\section{Introduction}

Studies in ecological genetics have documented temporal fluctuations in fitness among alleles in many species \citep{dobzhansky1943genetics,fisher1947spread,lynch1987consequences,cain1990population,cook1996medionigra,saccheri2008selection,bell2010fluctuating,bergland2014genomic,messer2016can} and long-term ecological field studies have documented fluctuations in the recruitment and death components of Malthusian parameters in many others \citep{caceres1997temporal,hoekstra2001strength,bell2010fluctuating,leigh2007neutral,kalyuzhny2014niche,kalyuzhny2014temporal,chisholm2014temporal}. However the extent to which this variability affects the long-term outcome of competition and biodiversity is not well understood. One obvious reason is that even long-term field studies offer but a fleeting glimpse into ecological and evolutionary processes that play out across millennia.

Theory partially fills this gap in knowledge by providing means to explore long-term outcomes. Typically, a dynamical model is proposed with parameters that have been calibrated empirically over relatively short time scales \citep{caceres1997temporal,usinowicz2012coexistence,chu2015large,ellner2016quantify,usinowicz2017temporal}. The model is then analyzed, numerically or analytically, to determine the conditions under which fluctuations in competition promote (or destabilize) biodiversity, to assess the impact that such fluctuations have on rates of allelic/species turnover and to evaluate the likelihood that new alleles/species will invade and become established in the population/community. From hereon ''species" in a ''community" is synonymous with ''alleles" in a ''haploid population". \emph{Mutatis mutandis}, such analyses are also relevant to game theory where different strategies are considered as alleles or species.

Most of the literature in the field, particularly a series of papers that laid the foundations of modern coexistence theory ~\citep{chesson1994multispecies,chesson2000mechanisms,chesson2003quantifying,HilleRisLambers2012,barabas2018chesson,grainger2019invasion}, focuses on invasibility, assuming that coexistence reflects the ability of each species to increase in frequency at low densities. Invasibility of a given species is then quantified using the mean logarithmic growth rate when rare $\mathbb{E}[r]$. Given a time series of (low) frequencies $\{x_t,x_{t+\Delta t},x_{t+2\Delta t}...\}$, $\mathbb{E}[r]$  is defined as~\citep{chesson2003quantifying},
\begin{equation}
\mathbb{E}[r] \equiv \frac{1}{\Delta t} \mathbb{E}\left[\ln \left(\frac{x_{t+\Delta t}}{x_t} \right) \right].
\end{equation}
When the number of individuals in the community, $N$, is large,
the \textbf{sign} of $\mathbb{E}[r]$ provides important information about invasibility and persistence properties ~\citep{chesson1982stabilizing,schreiber2011persistence,schreiber2012persistence,yahalom2019comprehensive}. If $\mathbb{E}[r]<0$ then the probability of invasion is negligible and the time to extinction depends logarithmically on the initial density of the focal species. If $\mathbb{E}[r]>0$ then the probability of invasion is larger than zero and the mean time to extinction depends on the size of the community, $N$.

While $\mathbb{E}[r]$ is a useful binary classifier, many authors have also assumed that its \textbf{magnitude} serves as a metric for persistence (i.e, ``a bigger $\mathbb{E}[r]$" implies greater invasibility and/or longer persistence times). Based on this assumption, several authors have attempted to parse the contributions of various ecological mechanisms to persistence through their contributions to the value of $\mathbb{E}[r]$~\citep{ellner2016quantify,letten2018species}. In other works, $\mathbb{E}[r]$-based metrics have been used to assess persistence in natural communities ~\citep{usinowicz2017temporal}.

This usage of $\mathbb{E}[r]$, and in particular its application when environmental stochasticity induces stable coexistence, is criticized in recent work by~\cite{pande2019mean} who showed that the magnitude of $\mathbb{E}[r]$ is much less informative than its sign (some of their arguments are reproduced in Appendix \ref{newsect}). Invasion probabilities and times to extinction are not single valued functions of $\mathbb{E}[r]$. Consequently, different combinations of the underlying parameters may yield the same $\mathbb{E}[r]$ yet with varying implications for invasion and persistence. Worse, invasion probabilities and times to extinction are not monotonic functions of $\mathbb{E}[r]$ and in some cases larger $\mathbb{E}[r]$s decrease invasion probabilities and times to extinction.

The aim of this paper is to suggest an alternative framework for community dynamics in fluctuating environments. We shall focus on the capacity of fluctuating environments to promote coexistence (stochasticity-induced stabilization or SIS). In our new framework we analyze abundance variations along the \textbf{arithmetic abundance} (or frequency) axis, as opposed to the log-abundance approach which is natural when the analysis is based on $\mathbb{E}[r]$. A new parameter, the ratio between the expected arithmetic growth when rare, $\mu$, and its variance, $g$, is shown to be more informative than $\mathbb{E}[r]$.

Our method has two limitations. First, we consider only one dimensional cases, like two species dynamics or multispecies communities in which a single species may be examined under the ``effective field" of all others. Second, we limit our discussion to the parameter regime in which abundance variations are not large and the diffusion approximation~\citep{crow1970introduction,karlin1981second} holds.

However, within its regime of validity our $\mu/g$ approach outperforms $\mathbb{E}[r]$-based methods.  $\mu/g$ yields the same threshold associated with the sign-change of $\mathbb{E}[r]$, so this parameter provides the same binary classification. In addition, both the probability of invasion and the time to extinction are single valued, monotonically increasing functions of $\mu/g$, so this parameter  provides a quantitative measure of invasibility and persistence which $\mathbb{E}[r]$ does not.

Another advantage of the new approach is its flexibility. To calculate $\mathbb{E}[r]$ one considers the dynamics of a rare species along the log-abundance axis. This choice is natural when the growth rate is a linear function of abundance ($dx/dt = r(t) x$) because the dynamics on the log-abundance scale correspond to a simple random walk. However this choice is unnatural when growth rates are not linearly dependent on abundance; examples include the Allee effect and recessive alleles where growth is a quadratic function of abundance ($dx/dt = r(t) x^2$). The $\mu/g$ analysis presented here is more transparent, more flexible, and allows for a synthesis of different cases (including quadratic dependencies) within a common mathematical framework.

To facilitate presentation, we restrict our discussion in two ways. First, we focus on large, well-mixed communities where the impacts of demographic stochasticity are weak and spatial structure can be neglected. Second, the amplitude of selection depends on the environmental conditions and, for simplicity, we allow the selection coefficients to adopt either of two values (dichotomous, or telegraphic, stochasticity) -- the generalization to any type of stochasticity being straightforward~\citep{yahalom2019comprehensive}.  We believe that these restrictions are irrelevant to the general qualitative and quantitative insights that emerge from the models considered below.

Our arithmetic scale approach and  the $\mu/g$ parameter are presented in detail in section \ref{sec2}. In section \ref{sec3} we implement this framework and explore the means by which fluctuating selection promotes coexistence in four models: the lottery model, the serial transfer model, a diploid genetic model with complete dominance, and the Moran/chemostat model. We focus on the threshold value for SIS using $\mu/g$ rather than the $\mathbb{E}[r]$; a detailed analysis of  the relationships between $\mu/g$ and $\mathbb{E}[r]$ is provided  in Appendix \ref{newsect}.

Section \ref{demographic} is devoted to the main advantage of the $\mu/g$ method -- its use as a quantitative measure of invasibility and stability. We show that both the probability of invasion ${\cal E}_+$ (Eq. \ref{invv}) and the mean persistence time $T$ (Eqs. \ref{extime} and \ref{extime1}) are single valued, monotonically increasing functions of $\mu/g$.  When the SIS mechanism fails, the system reaches extinction through very interesting dynamics that involve rejuvenation bursts, and this phenomenon is explored in Section \ref{extdyn}.

Section \ref{sec4} clarifies the relationships between SIS and certain life history strategies. Contrary to the prevailing view that adult longevity, seed banks, diapause etc. are essential to SIS, we show that one of their main effects is to reduce the effective amplitude of fluctuating selection which impedes SIS. Nevertheless, this may be counterbalanced by other features of the same life history strategies which improve SIS, such as increasing the effective generation time.

Section \ref{sec4} clarifies the relationships between SIS and certain life history strategies. Contrary to the prevailing view that adult longevity, seed banks, diapause etc. are essential to SIS, we show that one of their main effects, reducing the effective amplitude of fluctuating selection, impedes SIS. Nevertheless, other features of these same life history strategies, such as increasing the effective generation time, improve SIS. The net effect on SIS needs to be quantified on a case-by-case basis. Finally, we summarize our main insights and suggest new directions for future research in the discussion.

\section{$\mu/g$ analysis in the arithmetic frequency domain}\label{sec2}

Consider an arbitrary zero-sum two-species competition in an infinite community (no demographic stochasticity), with the focal species at frequency $x_t$ at time $t$. Let the dynamics of $x$ be described by a map that gives $x_{t+\delta}$ in terms of $x_t$ and other parameters, including the state of the environment $E$. Analyzing such maps is standard practice in population genetics, where $\delta$ is the generation time and the generations are non-overlapping. Here, we shall think of $\delta$ as the \emph{typical persistence time of the environment} [measured in units of one generation, see~\citep{danino2016effect}]. In what follows we provide examples in which this persistence time is shorter, longer and also equal to the generation time.

To understand stochasticity induced stabilization, consider a single step in this map from $x_0$ to $x_\delta$. We now omit the index $\delta$ and speak about $x_1$ as a function of $x_0$ and $E$, the prevailing environmental conditions. Accordingly,
\begin{equation} \label{map}
x_1 = \Phi(x_0,E).
\end{equation}
If the environment is fixed (i.e. if there is no environmental stochasticity) then $E$ is fixed and for every $m$ the map $x_{m+1} = \Phi(x_m,E)$ has the same parameter $E$. A simple map, familiar from population genetics, gives the focal allele frequency in the next haploid generation as
\begin{equation} \label{map_haldane}
\Phi(x_0,E) = \frac{x_0 W(E)}{x_0 W(E) + (1-x_0)},
\end{equation}
where $W(E)$ is relative fitness.

In a closed system (with no migration, no speciation/mutation, and with fixation and extinction as absorbing states) $\Phi$ must satisfy two conditions:
\begin{enumerate}
 \item $\Phi(0,E)=0$, meaning that the dynamics halts at $x_0=0$, the extinction point.
 \item $\Phi(1,E)=1$, so the dynamics halts at $x=1$, the fixation point.
 \setcounter{nameOfYourChoice}{\value{enumi}}
\end{enumerate}
We also require that,
\begin{enumerate}
\setcounter{enumi}{\value{nameOfYourChoice}}
\item For any \emph{fixed} $E$, either $\Phi(x,E) > x$ for $0<x<1$ or $\Phi(x,E) < x$ for $0<x<1$. This condition is \emph{sufficient} to ensure that under fixed environmental conditions the map (\ref{map}) has only two fixed points, at $x=0$ and $x=1$.
\setcounter{nameOfYourChoice}{\value{enumi}}
\end{enumerate}

In Figure \ref{fig1} (which refers to a specific example, given in Eq. (\ref{lot_map}) below) the red and the blue curves represent maps that satisfy conditions (1-3). Under the red map, $x_{m+1} < x_m$ and the frequency shrinks to zero. Under the blue map, $x_{m+1} > x_m$ and the frequency grows to one.

\begin{figure}
 \begin{center}
\centerline{\includegraphics[width=5.5cm]{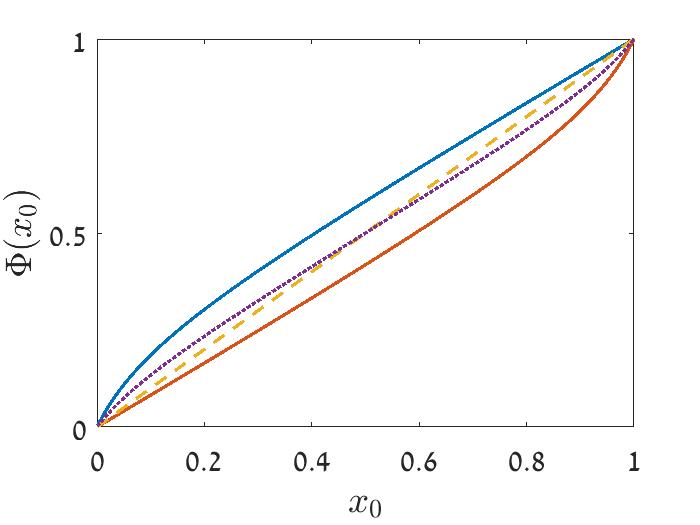}
\includegraphics[width=5.5cm]{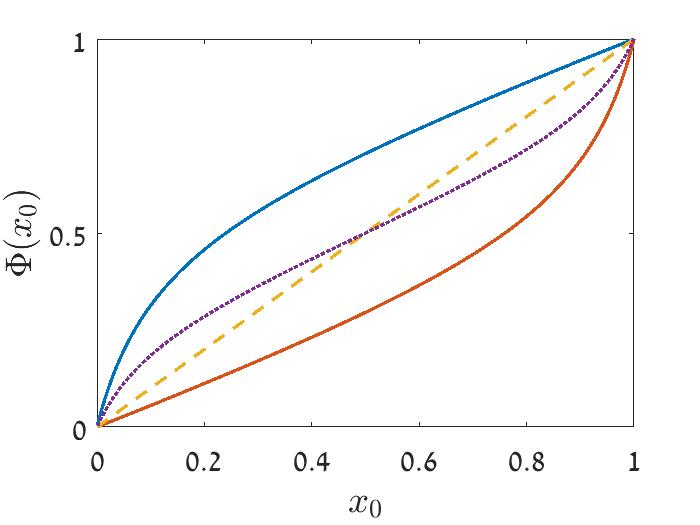}
\includegraphics[width=5.5cm]{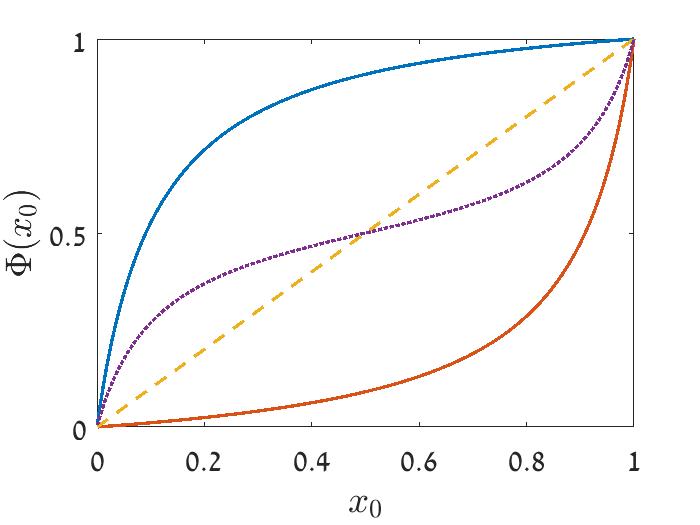}}
\end{center}
\caption{The lottery map, $\Phi(x_0,E)$ in Eq. (\ref{lot_map}), is plotted against $x_0$ for $\delta = 0.2$ (left), $\delta = 0.5$ (middle) and $\delta=1$ (right). Crossing of the dashed yellow line $x_1=x_0$ corresponds to the fixed points of the map where $\Phi(x_0,E) = x_0$. Above the yellow line the frequency of the focal species grows, below this line the frequency of the focal species shrinks. $\Phi(x_0,10)$ (blue line) always lies above the dashed yellow diagonal and so the focal species sweeps to fixation. $\Phi(x_0,0.1)$ (red line) always lies below the dashed yellow diagonal and so the focal species goes extinct. The mean map, $(\Phi(x_0,10)+\Phi(x_0,0.1))/2$ is represented by the dotted purple line. The expected growth of each species is positive when each is rare and the mean map has an attractive fixed point at $x_0=0.5$. Although the mean of the arithmetic growth increases as $\delta$ increases, SIS decreases with $\delta$, and vanish when $\delta=1$, because of the diffusive trapping, as demonstrated in Fig. \ref{fig2}}\label{fig1}
\end{figure}

What happens when the environment can change state? For example, let $x_{m+1} = \Phi(x_m,E_m)$ and give $E_m$ a stochastic process. It turns out that changes in $E_m$ promote coexistence, defined here as the ability of each species to invade the system when rare, if two criteria are fulfilled.

First, each species must have \textbf{positive expected growth when rare:} the arithmetic mean of abundance variations when $x \ll 1$ is positive.

Second, each species must avoid “\textbf{diffusive trapping when rare}”. As has long been known, a positive expected growth is \textbf{not} sufficient for invasion~\citep{lewontin1969population}. Under environmental stochasticity abundance variations are proportional to $x$, so the system spends a lot of time in the vicinity of $x=0$, the danger zone where demographic fluctuations may lead to extinction (alternatively, on the log-abundance axis the variations are $x$-independent, but the danger zone is now infinitely wide). Persistence is ensured only if the expected growth is sufficiently strong to overcome this diffusive trapping, for which the threshold condition is $\mathbb{E}[r]>0$.

Let us explain the conditions under which a map like (\ref{map}) allows for positive expected growth when rare and prevents diffusive trapping.

\subsection{Positive expected growth when rare}

We assume two species competition, The frequency of the focal species is $x$ and the frequency of its rival is $1-x$.  If $x \ll 1$, the map \ref{map} takes the form,
\begin{equation}
x_1 \approx \Phi(0,E)+x_0 \left. \frac{\partial \Phi(x,E)}{\partial x}\right|_{x=0}.
\end{equation}
Condition (1) above ensures that the first term vanishes and so the arithmetic mean abundance will grow on average if,
\begin{equation} \label{invasion}
\mathbb{E} \left[\left. \frac{\partial \Phi(x,E)}{\partial x}\right|_{x=0} \right] \equiv \mathbb{E} \left[\Phi'(0,E) \right] > 1
\end{equation}
where $\mathbb{E}$ is the expectation (average) taken over all possible environments $E$ (given their probabilities) for a fixed value of $x$. On the other hand, when $x_0$ is close to one, condition (2) implies that $\Phi(1,E)=1$ so $x_1 \approx 1+(x_0-1) \Phi'(1,E)$ and $x_1 -x_0 \approx (1-x_0)[1- \Phi'(1,E)]$. Accordingly, as long as $ \mathbb{E} \left[\Phi'(1,E)\right]>1$ the arithmetic mean growth of the focal species is negative close to the fixation point. To conclude, the conditions for the averaged map to produce positive arithmetic growth for either species when rare are,
\begin{equation}\label{bothconditions}
\mathbb{E} [\Phi'(0,E)]>1 \quad {\rm and} \quad  \mathbb{E}[\Phi'(1,E)]>1.
\end{equation}

\subsection{Avoiding diffusive trapping when rare}

Even with $\mathbb{E}[\Phi'(0,E)]>1$ invasion is not guaranteed. This is because the dynamics slow dramatically near the extinction point, a phenomenon we call diffusive trapping. To understand diffusive trapping we again focus on the frequency dynamics when $x \ll 1$,
\begin{equation} \label{delta}
\Delta(x) = x_1 - x_0 \approx x_0 (\Phi'(0,E) -1).
\end{equation}
Now $\Phi'(0,E) -1$ is the instantaneous expected growth of the focal species when rare, and at any moment is either positive or negative. For sufficiently small changes in frequency, $\Delta(x)$, one may use a diffusion approximation~\citep{karlin1981second}; the probability $P(x,t)$ (of finding the focal species at frequency $x$ at time $t$) then satisfies the Fokker-Planck equation (also known as the Kolmogorov forwards equation),
\begin{equation} \label{kolmogorov}
\frac{\partial P(x,t)}{\partial t} = \frac{\partial^2 }{ \partial x^2} \left( \frac{Var[\Delta(x)]}{2} P(x,t) \right) - \frac{\partial }{ \partial x} \left( \mathbb{E}[\Delta(x)] P(x,t) \right).
\end{equation}
 Given (\ref{delta}), when $x$ is very small (close to the extinction point) this equation takes the form
\begin{equation} \label{pp}
\frac{\partial P(x,t)}{\partial t} = \frac{\partial^2 }{ \partial x^2} \left[ g x^2 P(x,t) \right] - \frac{\partial }{ \partial x} \left[ \mu x P(x,t) \right],
\end{equation}
where [plugging Eq. (\ref{delta}) into the definition in Eq. (\ref{kolmogorov})],
\begin{equation}
\mu = \mathbb{E}[\Phi'(0,E) -1],
\end{equation} and
\begin{equation}
g = \frac{Var[\Phi'(0,E) -1]}{2}.
\end{equation}

If $P(x)$ reaches an equilibrium value $P_{eq}$, such that
$dP_{eq}/dt =0$, then as $x \to 0$ Eq. (\ref{pp}) implies,
\begin{equation} \label{condition}
P_{eq}(x \to 0) \sim C x^{\frac{\mu}{g}-2},
\end{equation}
where $C$ is a normalization constant.

Coexistence means that the equilibrium probability distribution can be normalized~\citep{schreiber2012persistence} i.e.,
\begin{equation}\label{binarycondition}
\frac{\mu}{g}=\frac{2 \mathbb{E}[\Phi'(0,E) -1] }{Var[\Phi'(0,E) -1]}>1.
\end{equation}
This condition prevents $P(x)$ in Eq. (\ref{condition}) from diverging in the extinction zone faster than, or equal to, $x^{-1}$. If for both species (each analyzed separately when rare) $\mu/g>1$,  then $P_{eq}(x)$ is normalizable. Otherwise $\mu/g \leq 1$ and $P_{eq}$ diverges in either or both extinction zones so fast that the equilibrium probability distribution cannot be normalized and so the system does not support coexistence. When $P_{eq}(x)$ is not normalizable in a fixed environment, yet is normalizable in a randomly fluctuating environment, the system supports SIS.

As explained in Appendix \ref{newsect}, as long as the diffusion approximation holds, the binary condition, $\mu/g>1$ (Eq. \ref{binarycondition} ), is equivalent to the condition $\mathbb{E}[r]>0$, where $\mathbb{E}[r]$ is the mean logarithmic growth rate when rare. Unlike $\mathbb{E}[r]$, the numerical value of $\mu/g$ is related to the probability of invasion (Eq. \ref{invv}) and also to the persistence time of the system (Eqs. \ref{extime} and \ref{extime1}).

\section{SIS in discrete time maps} \label{sec3}

In this section we explore four examples of SIS in discrete maps and show how they can all be analyzed in a common framework using $\mu/g$. We identify the threshold parameter above which stochasticity promotes coexistence in these systems. In three of these examples the traditional approach using logarithmic-abundance analysis and the condition $\mathbb{E}[r]>0$ can be implemented, and it yields the same threshold. The exception is the quadratic growth example in \ref{diploid}, where the traditional analysis finds $\mathbb{E}[r]=0$ while our $\mu/g$-based analysis predicts invasion. This section sets the scene for the quantitative analysis of invasibility and persistence (Section \ref{demographic}), for which the $\mu/g$ approach is necessary.

\subsection{The lottery model}

In the lottery model \citep{chesson1981environmental} a fraction $\delta<1$ of the population dies and is replaced according to probabilities that reflect the species abundances weighted by their fitnesses.  For the sake of concreteness, we consider a forest with only two tree species, where each dead adult tree leave an open gap. The chance of each species to recruit a gap is proportional to is relative seed (or larvae) frequency, and a focal species individual  produces $E$ seeds per each seed produced by its rival species.

 As the probability of an individual dying at each step is $\delta$, so lifetimes are distributed geometrically with mean $1/\delta$. $E$ is picked independently at random at each step and so $\delta$, as defined here, is the correlation time of the environmental variation when measured in units of a generation (unlike \cite{chesson1981environmental} who extended their theory to the case where $\delta$ is an arbitrary time step so that the environment may remain constant over multiple $\delta$s).

The chance that the focal species replaces a dead adult tree is proportional to its seed fraction,
$ E x_0/[E x_0 + (1-x_0)]$, and the death toll  is $\delta x_0$. Therefore,
\begin{equation} \label{lot_map}
x_1 = \Phi(x_0,E) = x_0 \left(1 - \delta + \delta \frac{ E }{E x_0+(1-x_0)} \right)
\end{equation}
Note that $\Phi$ satisfies all the conditions (1-3) above: $x_1=0$ if $x_0=0$, $x_1=1$ if $x_0=1$, $x_1>x_0$ if $E>1$, while $x_1<x_0$ if $E<1$.

Conditions \ref{bothconditions} for the arithmetic growth at rarity are translated to,
\begin{eqnarray}
\mathbb{E} \left[\Phi'(0,E)\right]=\mathbb{E}[1+(E-1)\delta]>1 \quad &\Rightarrow& \mathbb{E}[E]>1 \nonumber \\
\mathbb{E} \left[\Phi'(1,E)\right]=\mathbb{E} [1-\frac{(E-1)}{E}\delta]>1 \ \ \  &\Rightarrow& \mathbb{E}[1/E]>1.
\end{eqnarray}
If $E=1/2$ with probability $1/2$ and $E=2$ with probability $1/2$, then $\mathbb{E}[E]=\mathbb{E}[1/E] = 1.25>1$ and, with both conditions satisfied, the average map favors SIS. Several more examples, with $E$ jumping between $0.2$ and $5$ and between $0.1$ and $10$ are illustrated in Figure \ref{fig1}.

Figure \ref{fig1} shows that arithmetic growth when rare increases as $\delta$ increases. Yet SIS does not strengthen. Indeed, just the opposite is true -- the higher the value of $\delta$, the weaker is SIS. In our arithmetic-abundance framework this behavior is attributed to  diffusive trapping.

Set $E = e^s$, so that $E=1$ corresponds to $s=0$, and assume that $s \ll 1$. Accordingly, $\Phi'(0,E)-1 = \delta(e^s-1) \approx \delta (s+s^2/2)$. For a two state system where $s$ flips between $+\sigma$ and $-\sigma$, the first nonvanishing terms for the expected mean and expected variance of $\Phi'(0,E)-1$ are,
\begin{eqnarray} \label{ablot}
\mu &\approx& \frac{\delta\sigma^2}{2}\nonumber\\
g &\approx& \frac{\delta^2\sigma^2}{2}.
\end{eqnarray}
so $\mu/g=1/\delta$.
For small $x$,
\begin{equation} \label{chessoneq}
P_{eq} (x \to 0) \sim C x^{\frac{1}{\delta} -2}.
\end{equation}
Accordingly, SIS promotes coexistence ($P_{eq}$ is normalizable) as long as $\delta<1$~\citep{hatfield1989diffusion}. The probability of finding the system in the vicinity of the extinction point becomes larger and larger as $\delta$ increases. When $\delta =1$ the probability distribution function can no longer be normalized, meaning that the stabilizing effect has been killed (see Section \ref{extdyn}).

\begin{figure}
\begin{center}
\centerline{\includegraphics[width=5.5cm]{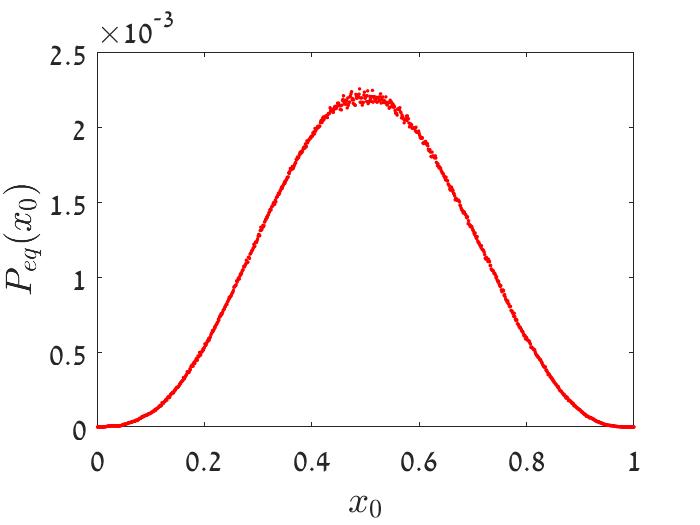}
\includegraphics[width=5.5cm]{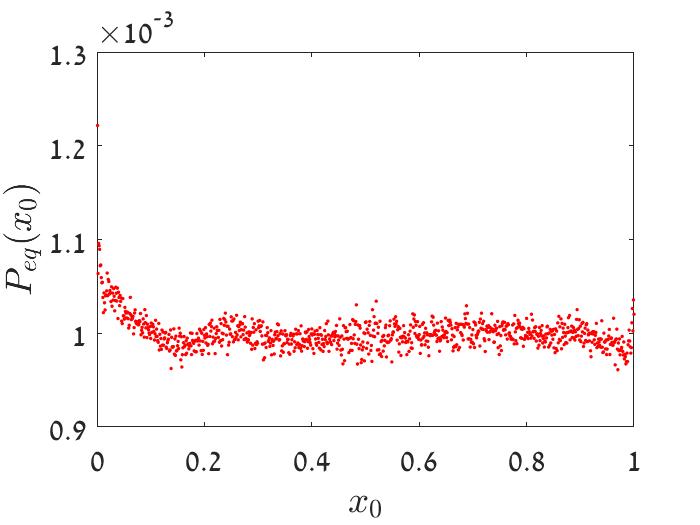}
\includegraphics[width=5.5cm]{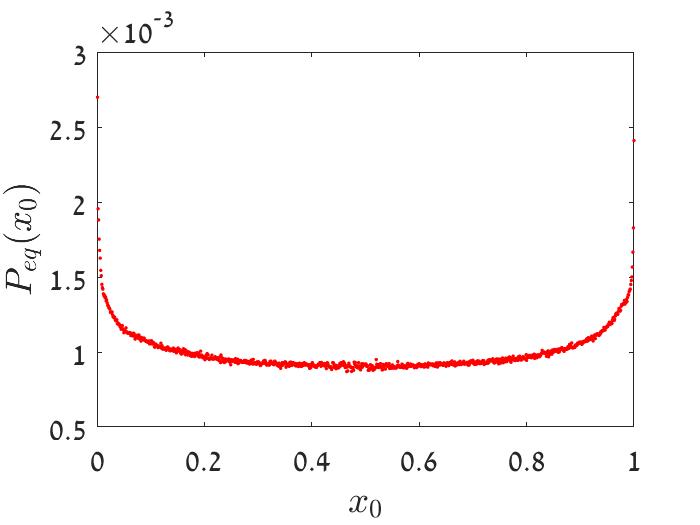}}
\end{center}
\caption{$P_{eq}$ as a function of $x$, as obtained from numerical simulation of Chesson's lottery map, $x_{t+1} = x_t [1-\delta+\delta E/(Ex+(1-x)]$, when $\delta = 0.2$ (left), $\delta = 0.5$ (middle) and $\delta = 0.54$ (right). The value of $E$ is picked at random in each step to be either $e^{0.3}$ or $e^{-0.3}$, the process was iterated $10^7$ times, starting from $x=1/2$. Plotted here are histograms of the number of visits in each bin (bin size 0.001) between zero and one. As predicted, the distribution is concave for $\delta < 0.5$, and convex above this value (we have chosen $\delta =0.54$ since for even higher $\delta$s the distribution peaks at zero and one are too high and blur the features of the distribution). When $\delta =0.5$ the predicted probability distribution is flat. In that case the outcome of the numerical experiment is, of course, more noisy but still indicates an $x$-independent distribution.} \label{fig2}
\end{figure}

Figure \ref{fig2} shows that the stabilizing effect indeed weakens as $\delta$ becomes larger and larger. The equilibrium distribution is concave when $\delta < 1/2$ and the system rarely approaches zero and one. The equilibrium distribution is convex when $\delta > 1/2$ with peaks at zero and one. While these peaks can be normalized if $\delta<1$, they nevertheless imply that the system spends lots of time close to the extinction zones. We shall discuss the quantitative outcomes in Section \ref{demographic}.

In Appendix \ref{newsect} we show, using the diffusion approximation, that  $\mathbb{E}[r] \sim \mu-g$.  When this approximation holds the condition $\mu/g=1/\delta>1$ implies $\mathbb{E}[r]>0$, and so the threshold for invasion is the same. When the diffusion approximation does not hold, our $\mu/g$ criteria fails and the threshold value is taken as $\mathbb{E}[r]>0$  [see~\cite{hatfield1989diffusion}]. On the other hand when the diffusion approximation holds, an increase in $\mu/g$ implies both a higher probability of invasion (Eq. \ref{invv}) and greater persistence (see Eqs. \ref{extime} and \ref{extime1}), while an increase in $\mathbb{E}[r]$ implies neither~\citep{pande2019mean}.

\subsection{The serial transfer regime} \label{dean}

Consider a single microbial species growing exponentially for time $T$ with growth rate $r$ until the carrying capacity of $N$ individuals is reached. After this growth phase the population is diluted by a factor $f$ into fresh medium and the process repeated. If $n_0 = f N$ denotes the population size immediately after the transfer, then the iterative map is,
\begin{equation} \label{dean1}
 n_1 = n_0 e^{r T}.
\end{equation}

Now consider two competing species. Starting from $n_A$ and $n_a = n_0-n_A$ individuals, each population grows exponentially (at rates $r_A$ and $r_a$) until the \emph{total} population reaches its carrying capacity of $N$ individuals. The community is then diluted into fresh medium and regrown. Note that the dilution step never alters the ratio of the competitors, $n_A:n_a$. Given $n_A$ and $n_a$, the time $T$ until the total population reaches $N$ is determined by,
\begin{equation}
n_A e^{r_A T} + n_a e^{r_a T} = N.
\end{equation}
With $n_0 = n_A + n_a = f N$ and setting $x_0 = n_A/n_0$ and $(1-x_0) = n_a/n_0$ yields,
 \begin{equation} \label{hard}
f x_0 e^{r_A T} + f (1-x_0) e^{r_a T} = 1.
\end{equation}
Once $T$ is found, the discrete time map for $x$ is,
\begin{equation}\label{hard2}
x_1 = x_0 f e^{r_A T}.
\end{equation}
This procedure can be iterated numerically: given $x_0$, $f$, $r_A$ and $r_a$, the transcendental equation (\ref{hard}) can be solved for $T$ which is then plugged into Eq. (\ref{hard2}) to yield $x_1$ and so on. Species $A$ wins as long as $r_A>r_a$ ($x_{i+1} > x_{i}$ for every $x_i$) and loses as long as $r_A<r_a$ ($x_{i+1} < x_{i}$ for every $x_i$). Moreover if $x_0 =0$ then $x_1=0$ and if $x_0=1$ then $x_1=1$. The serial transfer regime therefore satisfies conditions (1-3).

We would like to study the dynamics of this system when the values of $r_A$ and $r_a$ fluctuate in time, and in particular when their ratio $R = r_A/r_a$ changes randomly. Although we cannot solve Eq. (\ref{hard}) in general, we can focus on the behavior of $x$ when $x \ll 1$ (i.e., we can study invasion by species $A$). The problem is symmetric so our analysis is relevant to invasion by species $a$ as well.

When $x_0 \approx 0$ and $ 1-x_0 \approx 1$, Eq. (\ref{hard}) implies $e^{r_a T}=1/f$ or $T = -\ln(f)/r_a$. Plugging that into Eq. (20) yields,
\begin{equation} \label{dean2}
 x_1 = \Phi(x_0,R) = x_0 f e^{-r_A/r_a\ln f} = x_0 e^{(1-R)\ln f},
\end{equation}
The dilution factor is smaller than one so $\ln f <0$. If $R$ is fixed in time and $R>1$ then $x$ invades, while if $R<1$ then $x$ cannot invade. The expected (arithmetic) growth of the focal species is positive when $R$ varies if,
\begin{equation}
\mathbb{E}[\Phi'(0,R)] = \mathbb{E}[e^{(1-R)\ln f}] >1.
\end{equation}

As before, this condition is satisfied if we assume that $R$ can take only two values, $R = \exp(\pm \sigma)$. The model is symmetric and so the expected growth close to fixation is negative if $\mathbb{E}[e^{(1/R-1)\ln f}]>1$. Again, if in half of the cases $R = \exp(\sigma)$ and in half of the cases $R = \exp(-\sigma)$, then both conditions hold and the averaged arithmetic growth when rare is positive for both species.

\begin{figure}
\begin{center}
\centerline{\includegraphics[width=5.5cm]{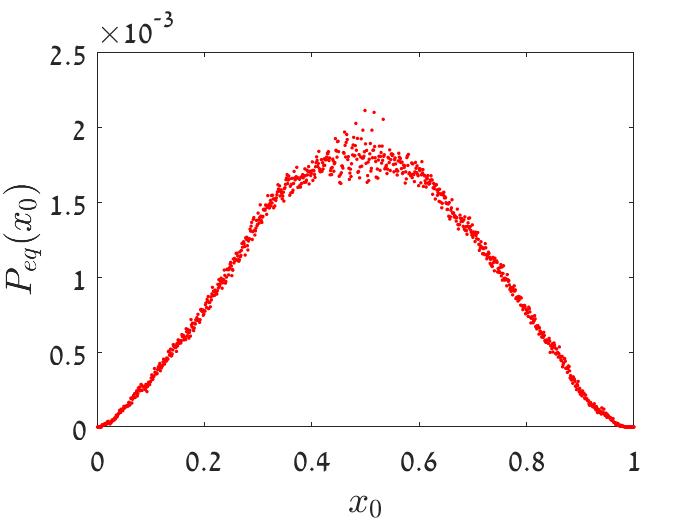}
\includegraphics[width=5.5cm]{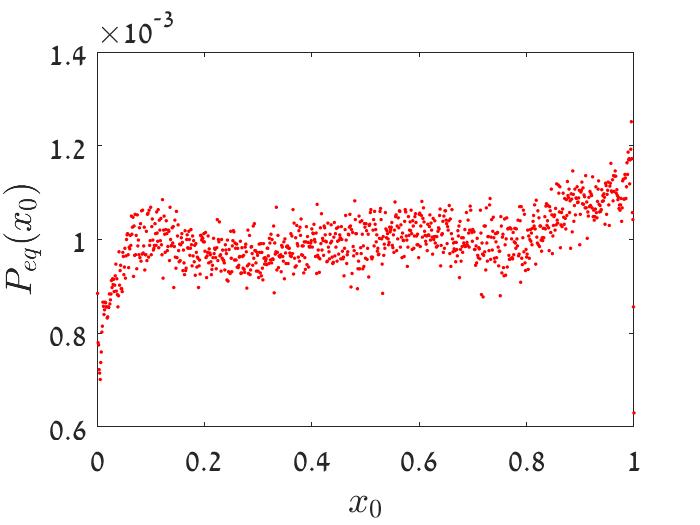}
\includegraphics[width=5.5cm]{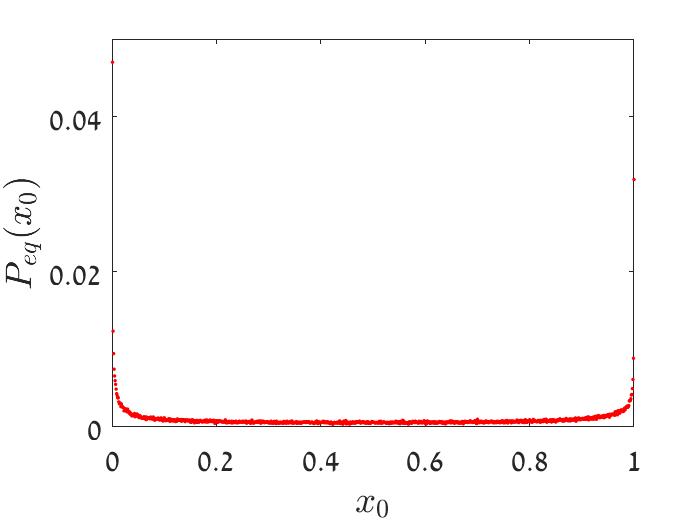}}
\end{center}
\caption{$P_{eq}$ as a function of $x$, as obtained from numerical simulation of the serial transfer map, for $R$ that may take two values, $R=1.1$ and $R=1/1.1$. The dilution factors are $f = 0.7$ (left panel) $f=1/e$ (middle panel) and $f=0.1$ (right panel). For each $x_0$ the value of $T$ was determined from Eq. (\ref{hard}) and $x$ was incremented according to (\ref{dean1}). Starting from $x=1/2$, the process was iterated 800000 times and the number of visits at each $10^{-3}$ bin was counted (see caption to Fig. \ref{fig2}). As expected fluctuations are more pronounced in the middle panel, when theory predicts a flat distribution.} \label{fig3}
\end{figure}

To determine the conditions needed to avoid diffusive trapping set $\sigma \ll 1$ so that,
\begin{equation}
\Phi'(0,R)-1 \approx e^{\pm \sigma \ln f - \frac{\sigma^2}{2} \ln f}-1 \approx \pm \sigma \ln f + \frac{\sigma^2}{2} (\ln f-1) \ln f.
\end{equation}
Now,
\begin{eqnarray}
\mu &=& \mathbb{E}[\Phi'(0,R)-1] \approx \frac{\sigma^2}{2}(\ln f-1)\ln f \nonumber \\
g &=& \frac{Var[\Phi'(0,R)-1]}{2} \approx \frac{\sigma^2}{2}(\ln f)^2,
\end{eqnarray}
and to the leading order $$\frac{\mu}{g} = 1 -\frac{1}{\ln f}.$$ Hence,
\begin{equation} \label{eq24}
P_{eq} (x \to 0) \sim C x^{-1-\frac{1}{\ln f}}.
\end{equation}
As $f<1$ so $\ln f<0$ and thus Eq. (\ref{eq24}) implies that the two species, $A$ and $a$, will coexist \textbf{at any level of dilution}. This means that SIS in the serial transfer model promotes coexistence even when the environmental correlation time far exceeds the doubling time. When $f = 1/e \approx 0.3679$ the power is zero and the probability distribution is uniform. For lower dilutions ($f>1/e$) $P_{eq}$ vanishes close to zero and this produces concave probability distributions. Higher dilutions, for which $P_{eq}$ diverges close to zero, produce convex distributions. These behaviors are illustrated in Figure \ref{fig3}.

\subsection{Diploids with dominance} \label{diploid}

\cite{haldane1963polymorphism} studied the dynamics of two alleles, with $A$ always dominant to $a$, in a randomly mating diploid population. If the fraction of $a$ alleles in the gamete pool is $x$ (and the fraction of $A$ is $1-x$) then, after random mating, the zygote genotypes follow classic Hardy-Weinberg proportions, with $AA: Aa: aa$ as $(1-x)^2: 2x(1-x): x^2$. Set the fitnesses of $AA$ and $Aa$ to one and the fitness of $aa$ to $W$. Then the frequency of $a$ alleles in the next generation's gamete pool is,
\begin{equation} \label{haldane}
x_1 = \Phi(x_0,W) = \frac{x_0(1+[W-1]x_0)}{1+(W-1)x_0^2}.
\end{equation}

Dominance introduces asymmetry to the model and so invasions by $a$ $(x \ll 1)$ and $A$ ($1-x \ll 1$) must be considered separately. Neither case is trivial.

Invasion by $a$ does not quite fit our prescribed framework, which assumes that $\Phi'(0,W)-1$ is either positive or negative, because map (\ref{haldane}) corresponds to the marginal case with $\Phi'(0,W)=1$ for any $W$. To find the conditions for invasion we need to take the next term in the series,
\begin{equation}
\Phi(x_0 \ll 1,W) \approx x_0 + (W-1)x_0^2.
\end{equation}
Clearly, invasion by $a$ requires $\mathbb{E}[W] >1$.

What about diffusive trapping? Given,
\begin{equation}
\Delta x = x_1-x_0 = (W-1) x_0^2,
\end{equation}
the relevant Kolmogorov forwards diffusion equation at small $x$ is,
\begin{equation} \label{pph1}
\frac{\partial P(x,t)}{\partial t} = \frac{\partial^2 }{ \partial x^2} \left( g x^4 P(x,t) \right) - \frac{\partial }{ \partial x} \left( \mu x^2 P(x,t) \right),
\end{equation}
where $g \equiv Var[W -1]/2$ and $\mu = \mathbb{E}[W -1]$. If $W= \exp(\pm \sigma)$ and $\sigma$ is small then both $g \approx \sigma^2/2$ and $ \mu \approx \sigma^2/2$. The $\sigma$ factors cancel and in the steady state (when the time derivative vanishes) the probability $P_{eq}(x)$ of finding the system at $x$ satisfies,
\begin{equation} \label{pph2}
\frac{\partial^2 }{ \partial x^2} \left( x^4 P_{eq}(x) \right) = \frac{\partial }{ \partial x} \left( x^2 P_{eq}(x) \right).
\end{equation}
Eq. (\ref{pph2}) has to be integrated twice. The constant of the first integration is dropped as it yields a term that diverges like $1/x^3$ close to zero. The resulting first order equation is exact and its solution is,
\begin{equation} \label{pph3}
P_{eq}(x \to 0) \sim C \frac{e^{-1/x}}{x^4}.
\end{equation}
The exponential term vanishes rapidly (faster than any power) as $x$ approaches zero and so the system cannot approach the extinction zone near $x=0$. Allele $a$ cannot go extinct.

Allele $a$ is a "super persistent" (although slow) invader for which diffusive trapping is strongly prohibited. This happens because only $aa$ homozygotes have fitness $W$, so the growth/decay rate of $a$ is a quadratic function of their abundance. When $A$ is common the $aa$ homozygotes are exceedingly rare and so, with most genotypes either $AA$ homozygotes or $Aa$ heterozygotes and having the same fitness of one, diffusive trapping is inefficient.

\begin{figure}
\begin{center}
\centerline{\includegraphics[width=5.5cm]{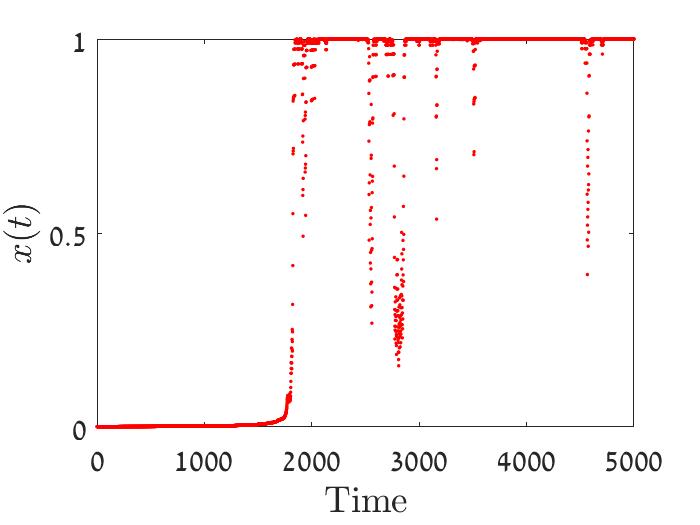}
\includegraphics[width=5.5cm]{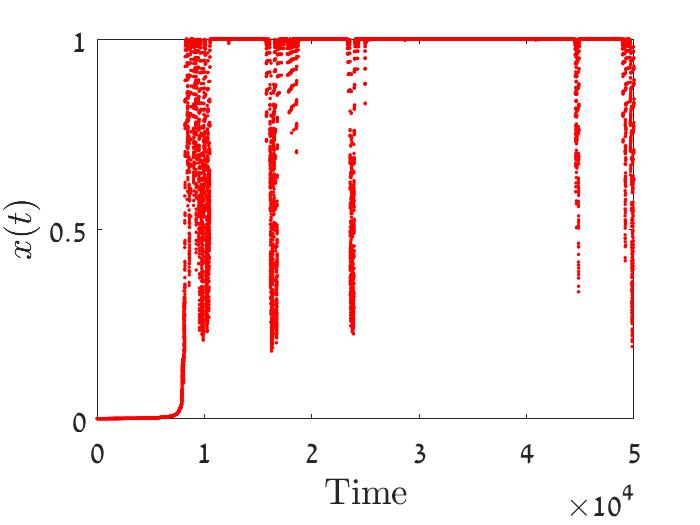}
\includegraphics[width=5.5cm]{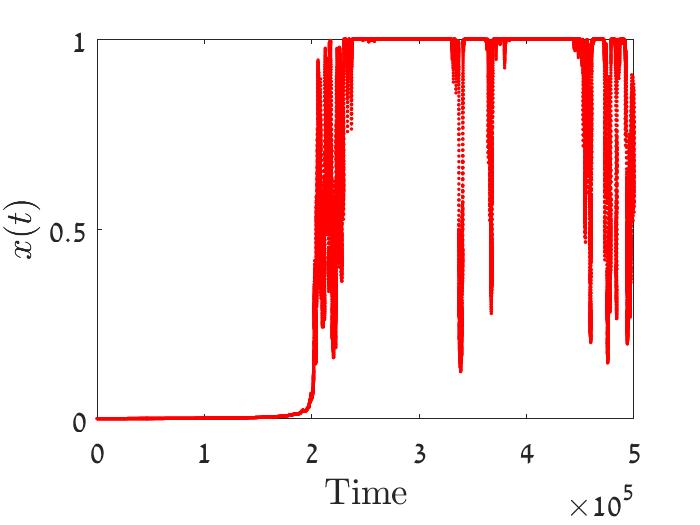}}
\end{center}
\caption{Typical trajectories of $x(t)$ as a function of $t$ (number of generations) for Haldane-Jayakar map, Eq. (\ref{haldane}) with $W = \exp(\pm \sigma)$, $\sigma =1$ (left) $\sigma =0.5$ (middle) and $\sigma = 0.1$ (right). The initial condition is $x_0 = 0.001$.
Nevertheless, all trajectories approach $x=1$ where they remain most of the time (the "bursts" of $A$ allele growth at long times are discussed in Section \ref{demographic}). While the expected growth of the rare species is positive at both ends, diffusive trapping nullifies this effect around $x=1$ and becomes inefficient close to zero. } \label{fig4}
\end{figure}

On the other hand, when $x$ is close to one,
\begin{equation}
\Phi'(1,W) = \frac{1}{W}
\end{equation}
If $W = \exp(\pm \sigma)$ then $\mathbb{E}[1/W]>1$, and the expected growth of the $A$ allele is positive when rare. However, growth by $A$ is linear in $(1-x)$. Once again $\mu=g=\sigma^2/2$ and the equilibrium distribution close to $x=1$ diverges as $(1-x)^{-1}$. As illustrated in Figure \ref{fig4}, the system is trapped: $A$ goes extinct and $a$ fixes~\citep{karlin1975random}.

\subsection{Models with asynchronous births and deaths} \label{cont}

The Moran and chemostat models describe populations with asynchronous births and deaths, with each birth obligately coupled to a death so that the size of the community, $N=n_A+n_a$, remains fixed. At each elementary  time step a species with abundance $n_A$ may stay at $n_A$, grow to $n_A+1$ or shrink to $n_A-1$.

We now have three time scales: 1) the time between elementary birth-death events, 2) the community generation time defined as $N$ birth-death events and 3) the correlation time of the environment, $\delta$, measured in units of a generation (e.g. if $N=10^6$ and $\delta = 0.01$, the correlation time of the environment is $10,000$ birth-death events).

Let us specify two microscopic versions of the Moran model. In both the log-fitness of a species, $s(t)$, is a stochastic process with mean 0 and variance $\sigma^2$.

The \textbf{local competition} model [model A of \cite{meyer2018noise}] describes the dynamics of two competing species where a chance encounter between two individuals ends up in a struggle over, say, a piece of food, a mate or a territory. To model that we pick, in each elementary birth-death event, two random individuals for a ''duel", the loser dies and the winner produces a single offspring. If the focal species is represented by $n_A$ individuals, then its fraction is $x = n_A/N$ and the chance for an interspecific duel is $2x(1-x)$. The probability the focal species wins a duel ($x\to x+1/N$) is $p = 1/2 + s(t)/4$ and the probability it loses ($x\to x-1/N$) is $1-p$. Accordingly, the change in the mean value of $x$ in a single birth-death event ($1/N$th of the generation time) is given by,
\begin{equation}
\mathbb{E}[dx/dt] =\mathbb{E}\left[\frac{x_1-x_0}{1/N}\right] = 2x(1-x)(2p-1) = s(t)x(1-x).
\end{equation}

In the \textbf{global competition} model ~\citep{moran1958random,dean2005protecting,dean2017fluctuating,dean2018haploids,meyer2018noise} an individual is chosen at random to die and then all the other individuals compete for the empty slot. The chance the focal species wins is proportional to its relative abundance weighted by its relative fitness; its chance to increase its abundance by one is $(1-x)xe^{s(t)}/[xe^{s(t)} + (1-x)]$. In such a case,
\begin{equation} \label{chesson}
\mathbb{E}[dx/dt] = \frac{x(1-x)(e^{s(t)}-1)}{ x e^{s(t)} + 1-x }.
\end{equation}

The two models are mathematically equivalent when there is no selection; setting $\sigma^2 =0$ defines the same purely demographic (neutral) process ($\overline{\Delta x}=0$) where the focal species has an equal chance, $x(1-x)$, to increase or decrease by one individual during each and every elementary birth-death step. If $s$ is fixed in time the two models differ slightly, but they both yield the same qualitative results: the focal species grows monotonically to fixation if $s>0$ and shrinks monotonically to extinction if $s<0$.

The two models behave very differently when the environment fluctuates. Let $s(t)$ take on either of two values, $\pm \sigma$. After each fixed period of time $\delta$ (after $N\delta$ elementary birth-death events) the system picks its state at random, either plus or minus. We need only analyze invasion by $x$ because the model is symmetric.

For the local competition model, when $x$ is small $\dot{x} = \pm \sigma x$ and so
\begin{equation}
x_1 = x_0 e^{\pm \sigma \delta}.
\end{equation}
This implies
\begin{equation}
\Phi'(0,\sigma)-1 \approx \pm \sigma \delta + \frac{\sigma^2 \delta^2}{2},
\end{equation}
assuming $\sigma \delta \ll 1$. Accordingly, $\mu=g=\sigma^2 \delta^2/2$ and once again diffusive trapping prevents either species invading. Stochasticity does not promote coexistence when competition is local.

In contrast, the global competition model~\citep{danino2016stability,meyer2018noise} yields,
\begin{equation} \label{eq4}
\mathbb{E}[dx/dt] \approx \pm \sigma x(1-x) +\sigma^2 x(1-x) (1/2-x)
\end{equation}
when $\sigma \ll 1$. It is the second term that makes the difference. When $x \ll 1$ the population size at time $\delta$ is,
\begin{equation}
x_1 = x_0 e^{\pm \sigma \delta +\sigma^2 \delta/2},
\end{equation}
so,
\begin{equation}
\Phi'(0,\sigma)-1 \approx \pm \sigma \delta + \frac{\sigma^2 \delta^2}{2} +\frac{\sigma^2 \delta}{2}.
\end{equation}
Accordingly,
\begin{eqnarray}
 \mu= \mathbb{E}[\Phi'(0,\sigma)-1] \approx \frac{\sigma^2 \delta^2}{2} +\frac{\sigma^2 \delta}{2} \nonumber \\
g = Var[\Phi'(0,\sigma)-1]/2 \approx \frac{\sigma^2 \delta^2}{2},
\end{eqnarray}
hence $$\mu/g \approx \frac{1}{\delta} +1,$$ and
\begin{equation} \label{ourmoran}
P_{eq}(x \to 0) \sim C x^{\frac{1}{\delta}-1}.
\end{equation}
With $\mu/g>1$ global competition allows the focal species to invade. By symmetry, its rival can also invade from rarity. Stochasticity promotes coexistence when competition is global. Like the serial transfer model, global competition promotes coexistence for any value of $\delta$, even when $\delta$ is larger than the generation time. In contrast, the lottery model becomes unstable when $\delta=1$.

\section{A quantitative analysis of coexistence} \label{demographic}

Until now we have classified systems into two groups, those that support SIS and those that do not, based on the behaviour of $P_{eq}(x)$ close to the extinction points $x=0$ and $x=1$. If $\mu/g>1$ then $P_{eq}(x)$ can be normalized and the system supports SIS. Otherwise $\mu/g\leq1$, the system is unstable and so $x$ becomes trapped in an extinction zone, either $x \ll 1$ or $1-x \ll 1$.

But what is the practical meaning of this? At the end of the day, biodiversity and polymorphism reflect a balance between the rates of invasion and the rates of extinction. In this section we implement the diffusion approximation and show that both the probability of invasion and the time to extinction are single-valued and monotonic functions of  $\mu/g$, by deriving explicit formulas for these quantities. Unlike the mean logarithmic growth rate when rare $\mathbb{E}[r]$ [see \citep{pande2019mean} and Appendix \ref{newsect}], $\mu/g$ is a proxy for the qualitative persistence properties of the system. This justifies the need for the new $\mu/g$-based synthesis presented in this paper.

\subsection{The strength of environmental variations and the approach to equilibrium}\label{appequ}

In all the cases considered so far, $P_{eq}(x \to 0)$ depends only on the correlation time $\delta$ and not on the amplitude of environmental variations, $\sigma$. Nevertheless, the rate at which the system approaches equilibrium is strongly affected by $\sigma$. For example, the forwards Kolmogorov equation for the lottery model, Eq. (\ref{pp}), has the form~\citep{hatfield1989diffusion,danino2016effect},
\begin{equation} \label{ppnew11}
\frac{\partial P(x,t)}{\partial t} = \sigma^2 \delta \left[ \delta \dfrac{\partial^2}{\partial x^2}\left(x^2(1-x)^2P(x,t)\right) - \dfrac{\partial}{\partial x}\left(x(1-x)(1-2x)P(x,t)\right) \right].
\end{equation}
So while $P_{eq}$ (determined on equating what is written in the square brackets to zero) is $\sigma$ independent, the time to equilibrium depends on $\sigma^2 \delta$. The smaller the $\sigma^2$, the longer the time to equilibrium. Clearly, when $\sigma^2=0$ there is no SIS and the system never reaches $P_{eq}$. Nor is SIS possible when $\sigma^2$ is so small that demographic stochasticity dominates selection ~\citep{danino2016effect,danino2016stability,hidalgo2017species,danino2018theory,meyer2018noise,dean2017fluctuating,dean2018haploids}.

The effects of decreases in $\delta$ are less obvious. On the one hand they slow convergence (the term $\sigma^2 \delta$ in Eq. (\ref{ppnew11}) becomes smaller). On the other hand, as $x \to 0$ so Eq. (\ref{ppnew11}) reduces to,
\begin{equation} \label{ppnew}
\frac{\partial P(x,t)}{\partial t} = \sigma^2 \delta \left[ \frac{\partial^2 }{ \partial x^2} \left( \delta x^2  P(x,t) \right) - \frac{\partial }{ \partial x} \left( x P(x,t) \right) \right].
\end{equation}
This is simply the lottery model version of Eq. (\ref{pp}). The equilibrium solution (Eq. (\ref{chessoneq})) is $P_{eq} = Cx^{1/\delta-2}$. Evidently, decreases in $\delta$ also reduce the chances of finding the system in an extinction zone near $x = 0$. In most cases the equilibrium properties dominate and decreasing $\delta$ strengthens SIS.

In what follows we will assume that the community is so very large that fluctuating selection dominates demographic stochasticity. This allows us to ignore all the effects of demographic stochasticity save one, namely that $x < 1/N$ means extinction.

\subsection{Invasion probabilities}

As shown in Appendix \ref{newsect}, for a system described by Eq. (\ref{pp})  the probability, $\mathcal{E}_+$, that a rare species (initial frequency $x_0$) invades a community (reaches some higher frequency $x^*$ before going extinct), is
\begin{equation} \label{invv}
\mathcal{E}_+ = \frac{1-e^{-\left(\frac{\mu}{g}-1\right) ln(x_0/{\epsilon})}}{1-e^{-\left(\frac{\mu}{g}-1\right) ln(x^*/{\epsilon})}},
\end{equation}
where $\epsilon$ is the frequency below which extinction is declared [naturally taken to be $x = 1/N$, see \cite{chesson1982stabilizing} and \cite{pande2019mean}]. For the lottery model we find (Appendix \ref{newsect}),
\begin{equation}
\mathcal{E}_+ = \frac{1-e^{-\left(\frac{2 \mathbb{E}[r]}{\delta^2 \sigma^2}\right) ln(x_0/{\epsilon})}}{1-e^{-\left(\frac{2 \mathbb{E}[r]}{\delta^2 \sigma^2} \right) ln(x^*/{\epsilon})}}.
\end{equation}
The probability of invasion depends not only on the mean logarithmic growth rate when rare, $\mathbb{E}[r]$, but also on the variance, $g = \delta^2\sigma^2/2$.

The sign of $\mu/g-1$ and the sign of $\mathbb{E}[r]$ are the same and a change in the sign of these quantities marks a qualitative change in behavior (see Appendix \ref{newsect}). A positive sign means SIS operates and the chance of invasion is $N$-independent for large $N$. A negative sign means SIS does not operate and any chance of invasion vanishes for sufficiently large $N$. For this binary decision - whether or not SIS is active - both $\mu/g$ and $\mathbb{E}[r]$ are decent parameters. However, above the transition point the chance of invasion is determined unequivocally by $\mu/g$, while the same $\mathbb{E}[r]$ may correspond to different values of $\mathcal{E}_+$ according to the value of $g$.

\subsection{Extinction rates and persistence times with SIS}

At any given time $t$ the rate of extinction is proportional to the chance that the system visits the extinction zone $x < 1/N$~\citep{chesson1982stabilizing,kessler2007extinction,meyer2018noise},

\begin{equation} \label{rate}
Rate \sim \int_0^{1/N} \ P(x,t) \ dx.
\end{equation}

For a \textbf{stable} system $\mu/g>1$ and, after some period of time, the probability distribution converges on $P_{eq} \sim C x^{\mu/g-2}$. With $N$ sufficiently large, the system "leaks" to extinction slowly through a tiny "hole" between zero and $1/N$. The rate of extinction (assuming factors not scaling with $N$ are negligible) is proportional to,

\begin{equation} \label{extime}
Rate \sim \int_0^{1/N} x^{\mu/g-2} \ dx \sim N^{-(\mu/g-1)}.
\end{equation}
When $\mu/g>1$, the rate of extinction decays to zero as $N$ goes to infinity.

For the lottery model one finds,
\begin{equation} \label{lotteryrate}
Rate \sim N^{-\left(\frac{2 \mathbb{E}[r]}{\delta^2 \sigma^2} \right)}
\end{equation}
and so extinction, like invasion, is not solely dependent on  $\mathbb{E}[r]$; the same $\mathbb{E}[r]$ may correspond to different $Rate$s according to the value of $g= \delta^2 \sigma^2/2$.

The mean time to extinction is inversely proportional to the rate of extinction, $T \sim N^{\mu/g-1}$. For all the models described in Section \ref{sec3} the time to extinction behaves like a power-law in $N$ (see \cite{yahalom2019comprehensive}). In particular, it has been shown for the Moran process \citep{hidalgo2017species,danino2016stability} that
\begin{equation} \label{extime1}
T \sim N^{\mu/g-1} = N^{1/\delta},
\end{equation}
in agreement with Eq. (\ref{ourmoran}).

\section{Extinction dynamics when SIS fails} \label{extdyn}

A different route to calculating persistence times must be used for \textbf{unstable} systems that lack SIS. With  $\mu/g\leq1$ the system never equilibrates and so the equilibrium probability distribution $P_{eq}$ does not exist.

To understand the extinction dynamics in such a system consider the simple, famous and solvable model,
\begin{equation}
x_{t+1} = \frac{E x_t}{E x_t+(1-x_t)}.
\end{equation}
This is the lottery model in the limit $\delta =1$. The quantity $u \equiv (1-x)/x$ satisfies $u_{t+1} = u_t/E$, and so $z = \ln u$ obeys the recurrence relation
\begin{equation} \label{eq47}
z_{t+1} = z_{t} - \ln E = z_t \pm \sigma,
\end{equation}
where the last identity holds for dichotomous noise with $E = \exp(\pm \sigma)$.

Accordingly, $z$ performs a simple random walk between $-\infty$ and $\infty$ with step size $\sigma$. The solution in the continuum limit is given by the diffusion equation with diffusion constant $\sigma^2$. If the initial condition is $x(t=0) =1/2$, then $u(t=0) = 1$ and $z(t=0) = 0$. The probability finding the system at $z$ at time $t$ is given by,
\begin{equation}
P(z,t) = \frac{e^{-z^2/2 \sigma^2 t}}{ \sqrt{2 \pi \sigma^2 t}}.
\end{equation}
Since $x = 1/(1+u)=1/(1+e^z)$, one finds \citep{gillespie1972effects},
\begin{equation} \label{infinite}
P(x,t) = \frac{e^{-\frac{\ln^2\left(\frac{1-x}{x}\right)}{2\sigma^2 t}}}{x(1-x) \sqrt{2 \pi \sigma^2 t}}.
\end{equation}

\begin{figure}
\begin{center}
\includegraphics[width=15cm]{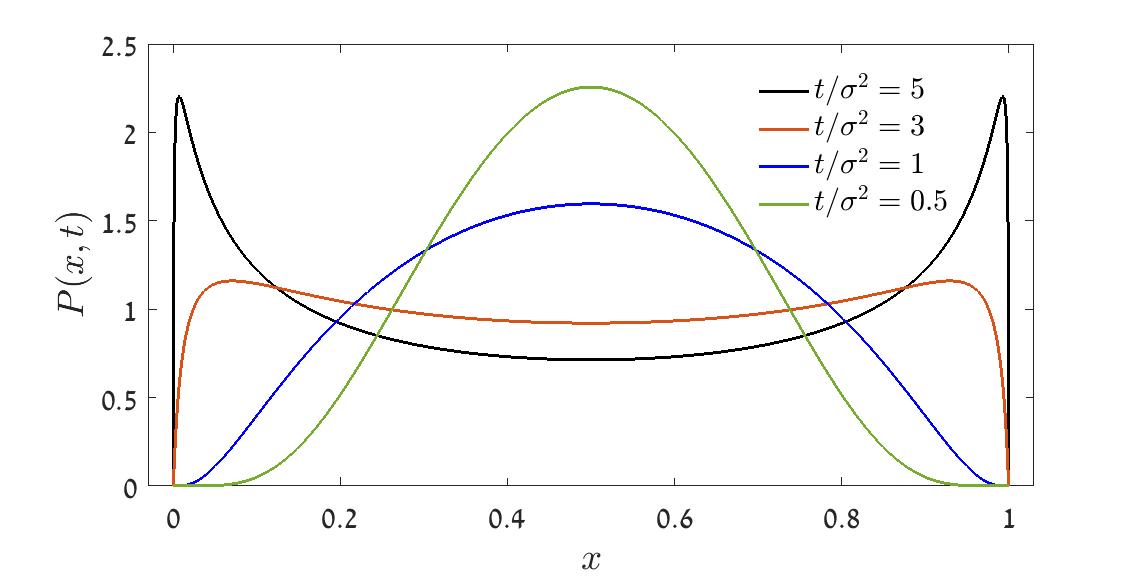}
\end{center}
\caption{The evolution of the probability distribution $P(x,t)$, Eq. (\ref{infinite}), through time. As $t/\sigma^2$ grows, the pdf develops wings at the extinction zone, still it drops sharply to zero when $x$ or $1-x$ are smaller than $\exp(-\sigma^2 t/2)$.} \label{fig5}
\end{figure}

Eq. (\ref{infinite}) allows us to understand diffusive trapping when SIS is inefficient. $P(x,t)$ is always a normalized and strictly legitimate probability distribution function that, with the passage of time, develops higher and higher "wings" in the extinction zones (Figure \ref{fig5}) as it attempts to reach its unattainable target, $P_{eq} = 1/x(1-x)$ [this formal solution to Eq. (\ref{ppnew11}) when $\delta =1$ is technically the \emph{infinite density} associated with Eq. (\ref{infinite}); see \citet{aghion2019non}]. The zone, $x<exp(-\sigma t/2)$, where the probability drops to zero is ever-shrinking and as time proceeds $x$ is found (on average) closer and closer to zero or to one. Since the step-size becomes vanishingly small when the frequency approaches zero, the system tends to spend more time in the extinction zone. At some stage the single individual limit at $x=1/N$ is crossed and the dynamics halts.

A rare sequence of good years may nevertheless cause a burst of growth in an extinction prone population. This phenomenon is clearly seen for the $A$ allele in Figure \ref{fig4}. With time the frequency of $A$ decreases and the corresponding probability density (which reflects an average over many specific histories) accumulates closer and closer to the extinction zone. Nevertheless, $A$ may rejuvenate given a long sequence of good years. As time proceeds, longer and longer sequences of good years are required to produce bursts, which become rarer and rarer, and also lower in amplitude, until the very last individual dies. On the other hand, right after a successful ''rejuvenative burst" the chance for another burst increases (shorter sequences of good years are needed) and so the bursts tend to cluster. The typical abundance time-series of an extinction prone allele is highly nontrivial, with an intricate burst structure that may provide false clues to long-term dynamics.

\section{Life histories, asymmetric competition, multispecies communities and SIS}\label{sec4}

The lottery model generates SIS only when generations overlap ($\delta<1$). In consequence, many authors have suggested that diversity is promoted by mechanisms ``for persisting during unfavourable periods, such as a seedbank, quiescence or diapause" ~\citep{kelly2014temporal}. The idea is that long-lived life-history stages shield individuals from competition thereby promoting diversity by buffering ``species against sudden rapid declines"~\citep{messer2016can}.

Our analysis suggests that sometimes this idea may be misleading. Experiments in the serial transfer regime~\citep{yi2013bounded} and analyses of global competition in the Moran/chemostat model~\citep{moran1958random,dean2005protecting,dean2017fluctuating,dean2018haploids,meyer2018noise} show that long-lived life-history stages shielding individuals from competition are not needed to promote coexistence. Nor are overlapping generations essential~\citep{abrams1984variability,dean2017fluctuating}. Continuous time exponential growth in the serial transfer model can always be replaced by discrete time non-overlapping generation geometric growth (1, 2, 4, 8 ... and 1, 3, 9, 27 ... etc.) and still yield SIS.

In general, mechanisms that reduce environmentally induced fitness fluctuations (i.e. reduce $\sigma^2$) diminish SIS. In the lottery model, $\mathbb{E}[r] = \sigma^2 (1-\delta)/2$ (see Eq. \ref{ER}, Appendix \ref{newsect}) disappears whenever a strategy, perhaps by an "ideal" bet-hedger, eliminates fitness variations. More quantitatively, the smaller the $\sigma^2$, the longer the time to the SIS associated equilibrium as explain in section \ref{appequ}. On the other hand, strategies like seed banks or diapause can strengthen SIS by increasing the effective generation time, hence decreasing $\delta$ (the correlation time of the environment, measured in units of a generation). Behaviors that weaken competition in unfavorable environments compared to favorable ones generate the subadditivity needed for long-term persistence~\citep{chesson1989short}.

Several other ecological processes weakening SIS are mentioned in the literature. Asymmetric competitions~\citep{danino2016stability,dean2017fluctuating,dean2018haploids,meyer2018noise}, where one competitor has higher mean fitness than the other, brings the equilibrium frequency of the inferior species closer to the extinction point. In this case the fitter competitor will almost certainly fix, although the time to fixation may still be very long. Increased community diversity also weakens SIS by reducing the mean frequency of each species~\citep{hatfield1997multispecies} thereby increasing the risk of stochastic extinction~\citep{danino2016effect}. Moreover, in a given environment fitter species increase in abundance so that the mean fitness of the community increases making more difficult for rare species to invade~\citep{danino2018theory}. All in all, the relationships between ecological processes and the capacity of SIS to promote diversity are far from trivial.

\section{Discussion}

In this paper we have provided a concise and intuitive introduction to the phenomenon of stochasticity-induced stabilization, illustrated with simple examples involving two species zero-sum competition dynamics with dichotomous (two-state) environmental fluctuations. We presented a new perspective on stabilization by fluctuating selection, one that focuses on the arithmetic mean growth when rare, $\mu$, and its variance, $g$, rather than the mean logarithmic growth rate when rare, $\mathbb{E}[r]$.

$\mathbb{E}[r]$ is an important binary indicator; when its sign is negative a species cannot invade from rarity, while if its sign is positive the chance of invasion is finite for an arbitrarily large $N$. Similarly, when $\mathbb{E}[r]>0$ for all species, the time to extinction diverges with $N$~\citep{schreiber2011persistence,schreiber2012persistence}. However, the magnitude of $\mathbb{E}[r]$ is not a good metric of persistence~\citep{pande2019mean} because the probability of invasion, ${\cal E}_+$, and the coexsistence time, $T$, are neither single-valued functions, nor monotonic functions, of $\mathbb{E}[r]$.

We have shown that, \emph{when the diffusion approximation holds,} the parameter $\mu/g$ is more informative than $\mathbb{E}[r]$. As a binary classifier $\mu/g>1$ is equivalent to $\mathbb{E}[r]>0$ (see Appendix \ref{newsect}). In addition, measures of persistence, such as the probability of invasion (Eq. \ref{invv}) and the time to extinction (Eqs. \ref{extime} and \ref{extime1}), are single-valued monotonically increasing functions of $\mu/g$. The $\mu/g$ method also has the added advantage of allowing us to analyze cases with nonlinear dynamics (e.g. a recessive allele where $dx/dt \approx E(t) x^2$ at low-frequency).

The $\mu/g$ method is naturally interpreted along the arithmetic-abundance (or arithmetic frequency) axis. The arithmetic amplitude of abundance variations (as opposed to their log-amplitude) shrinks to zero in the vicinity of $x=0$. As a result, if $\mu$ is not large enough the probability tends to accumulate in the low density region where demographic stochasticity may cause extinction (diffusive trapping). To achieve persistence, the effect of expected growth must overcome diffusive trapping.

Two elements that we did not not consider in this paper are the impacts of demographic stochasticity and what happens when the diffusion approximation fails. Fluctuating selection must be stronger in finite populations to maintain SIS in the face of demographic stochasticity. In models where the diffusion approximation holds this effect has been quantified~\citep{dean2017fluctuating,hidalgo2017species,danino2016stability,meyer2018noise,dean2018haploids}. Results for cases where the diffusion approximation fails are limited both in number and to models without demographic stochasticity~\citep{meyer2018noise}. Finding a metric analogous to $\mu/g$ when the diffusion approximation fails and in the presence of demographic stochasticity is an important question yet to be addressed.

\section{Acknowledgments}
Both authors contributed equally to the ideas and writing in this manuscript. N.M.S. acknowledges helpful discussions with Eli Barkai, David Kessler, Jayant Pande and Niv de-Malach. We thank Peter Chesson for several frank and thorough reviews which helped us greatly improve the manuscript. This research was supported by the ISF-NRF Singapore joint research program (grant number 2669/17).

\bibliography{refs}

\clearpage

\begin{center}
\Large{\bf{Appendix}}
\end{center}
\setcounter{equation}{0}
\setcounter{section}{0}
\setcounter{figure}{0}
\setcounter{table}{0}
\setcounter{page}{1}
\makeatletter
\renewcommand{\theequation}{A\arabic{equation}}
\renewcommand{\thefigure}{A\arabic{figure}}

\section{$\mathbb{E}[r]$ vs. $\mu/g$: a comparison} \label{newsect}

In this appendix we will clarify the relationships between our $\mu/g$ analysis and the usage of the mean logarithmic growth rate when rare, $\mathbb{E}[r]$, as a measure for invasibility and persistence. When a concrete example is needed we will consider the lottery model as defined in Eq. (\ref{lot_map}).

\subsection{Growth rate when rare}

First, we recite the definition of $\mathbb{E}[r]$, which is the mean growth rate along the log abundance (or log-frequency) axis. If for a certain species the outcome of two successive frequency measurements are $x_t$ and $x_{t+\Delta t}$, then by definition,
\begin{equation}
r = \frac{\ln (x_{t+\Delta t}/x_t)}{\Delta t}.
\end{equation}
Accordingly, given a time series of (low) frequencies $\{x_t,x_{t+\Delta t},x_{t+2\Delta t}...\}$ and so on, this mean growth rate is defined as~\citep{chesson2003quantifying},
\begin{equation} \label{pgr}
 \mathbb{E}[r] \equiv \frac{1}{\Delta t} \mathbb{E}\left[\ln \left(\frac{x_{t+\Delta t}}{x_t} \right) \right].
\end{equation}

For the discrete time lottery model Eq. (\ref{lot_map}) above (with $x_0 \ll 1$) implies:
\begin{equation} \label{ER}
 \mathbb{E}[r] \equiv \frac{\mathbb{E}[\ln(1-\delta + \delta E)]}{\delta}.
\end{equation}
When $E = e^{\pm \sigma}$ and the diffusion approximation holds,
\begin{equation} \label{ap9}
 \mathbb{E}[r] \equiv \frac{\ln(1-\delta + \delta e^\sigma)+\ln(1-\delta + \delta e^{-\sigma})}{2\delta} \approx \frac{\sigma^2}{2}(1-\delta).
\end{equation}
If $\mathbb{E}[r]$ is indeed a quantitative measure of persistence one should expect that probability of invasion and time to extinction grows with $\sigma^2$. As shown in detail in \citep{pande2019mean}, and as we shall see below, this is not the case. In the general case where the problem is asymmetric (see section 6) $\mathbb{E}[r]$ does not vanish with $\sigma^2$, but its usage as a qualitative indicator is still problematic~\citep{pande2019mean}.

\subsection{Fokker-Planck equation in log-abundance coordinates}

 Let us consider the Fokker-Planck equation for rare species, Eq. (\ref{pp}),
\begin{equation} \label{ap1}
\frac{\partial P(x,t)}{\partial t} = \frac{\partial^2 }{ \partial x^2} \left(g x^2 P(x,t) \right) - \frac{\partial }{ \partial x} \left( \mu x P(x,t) \right),
\end{equation}
To clarify the relationships between the stochastic process and $\mathbb{E}[r]$, we now switch to the log-frequency domain. To do that we substitute $$y=\ln x,$$
and use $Q(y,t)$ for the probability distribution function in terms of $y$. Since $P(x)dx = Q(y) dy$ one has, $$P(x,t) = Q(y,t)/x.$$

Measuring time in units of a single generation (such that $Q(y,t+\delta) \approx Q(y,t) + \delta \partial Q(y,t)/\partial t$) allows Eq. (\ref{ap1}) to be recast as,
\begin{equation}\label{mct7}
\frac{\partial Q(y,t)}{\partial t} = \frac{g}{\delta} \frac{\partial^2 Q(y,t)}{\partial y^2} - \frac{(\mu-g)}{\delta} \frac{\partial Q(y,t)}{\partial y }.
\end{equation}

For the lottery model we discovered above (Eq. \ref{ablot}) that $\mu = \delta \sigma^2/2$ and $g = \delta^2 \sigma^2/2$. A comparison with Eq. (\ref{ap9}) yields, for the lottery model, $\mathbb{E}[r]=(\mu-g)/\delta$ and so,

\begin{equation}\label{mct22}
 \frac{\partial Q(y,t)}{\partial t} = \frac{\delta \sigma^2}{2} \frac{\partial^2 Q(y,t)}{\partial y^2 } - \mathbb{E}[r] \frac{\partial Q(y,t)}{\partial y }.
\end{equation}

This is a general feature: as $\delta>0$, the coefficient of the first derivative term of the Fokker-Planck equation in logarithmic coordinates, which is (by definition) the mean growth rate when rare $\mathbb{E}[r]$, has the same sign as $\mu-g$ or $\mu/g-1$. Accordingly, when the diffusion approximation holds the threshold condition $\mathbb{E}[r]>0$ is  equivalent to $\mu/g>1$.

\subsection{The chance of invasion}

Eqs.~(\ref{mct7}) and (\ref{mct22}) are standard convection-diffusion equations for which many results are known. For example, it is known~\citep[Eq. 2.3.8]{redner2001guide} that if a particle starts at a distance $y_0$ above the extinction point, and if it invades only when it reaches $y^*$ before reaching zero, then the chance of invasion, $\mathcal{E}_+$, is,
\begin{equation} \label{invv1}
\mathcal{E}_+ = \frac{1-e^{-\left(\frac{\mu}{g}-1\right) y_0}}{1-e^{-\left(\frac{\mu}{g}-1\right) y^*}} = \frac{1-e^{-\left(\frac{2 \mathbb{E}[r]}{\sigma^2 \delta}\right) y_0}}{1-e^{-\left(\frac{2 \mathbb{E}[r]}{\sigma^2 \delta} \right) y^*}}.
\end{equation}
The expression on the far right is for the lottery model. In an invasion experiment, if the initial condition is $x=2\epsilon$ and invasion is declared when $x$ reaches $x_f$ before reaching $\epsilon$, then $y_0 = \ln 2$ and $y^* = \ln(x_f/\epsilon)$. If $\mu/g<1$ (or $\mathbb{E}[r]<0$) the exponential term in the denominator diverges when $N \to \infty$ ($\epsilon \to 0$ so $y^* \to \infty$ while $y_0$ is kept fixed). On the other hand when $\mu/g>1$ (or $\mathbb{E}[r]>0$) the exponential term vanishes so the chance of invasion is finite.

Accordingly, one observes that $\mu/g-1$ determines the probability of invasion from rarity. $\mathbb{E}[r]$ is merely the ``convection" term (the first derivative, which determines the mean velocity along the logarithmic axis in Eqs. (\ref{mct7}) and (\ref{mct22})); its value is not related directly to the probability of invasion because it does not account for the diffusion (second derivative) term. Only the \textbf{ratio} between convection and diffusion, $\mu/g-1$, determines the chance of invasion.

In particular, for the lottery model in this parameter regime the actual chance of invasion is independent of $\sigma^2$. When $\sigma^2$ increases while all other parameters are kept fixed, the ratio $\mu/g$ does not change. An increase in $\sigma^2$ leads to an increase in $\mathbb{E}[r]$, but this has nothing to do with invasibility.

A similar problem appears when extinction times are analyzed. As shown in section \ref{demographic}, this time scales like $N^{\mu/g}$, so when the diffusion approximation holds it is independent of $\sigma^2$, although the numerical value of $\mathbb{E}[r]$ increases with $\sigma^2$.

\end{document}